\documentclass[]{aa}

\usepackage[utf8]{inputenc}
\usepackage{float}
\usepackage{natbib}
\usepackage{graphicx}
\usepackage{subcaption}
\usepackage{numprint}
\usepackage{color}
\usepackage{txfonts}
\usepackage{hyperref}
\usepackage{tabularx}
\usepackage{amsmath}
\usepackage{amssymb}
\usepackage{siunitx}
\usepackage{mathtools}
\usepackage[normalem]{ulem}
\usepackage{tikz}
\usetikzlibrary{calc,shapes,arrows}
\usepackage{multirow}
\usepackage{threeparttable}
\providecommand{\Unit}[1]{\ensuremath{\mathrm{~#1}}} 

\providecommand{\parallax}{\ensuremath{\varpi}}

\newcommand\secref[1]{Sect.~\ref{#1}}
\newcommand\secrefalt[1]{Section~\ref{#1}}
\newcommand\figref[1]{Fig.~\ref{#1}}

\newcommand\figrefalt[1]{Figure~\ref{#1}}

\newcommand\tabref[1]{Table~\ref{#1}}
\begin{document}

\title{Hunting for open clusters in \textit{Gaia} DR2: $582$ new OCs in the Galactic disc\thanks{Tables 1 and 2 are only available in electronic form
at the CDS via anonymous ftp to cdsarc.u-strasbg.fr (130.79.128.5) or via http://cdsweb.u-strasbg.fr/cgi-bin/qcat?J/A+A/}}

\author{
    A. Castro-Ginard          \inst{\ref{inst:UB}}\relax
\and  C. Jordi          \inst{\ref{inst:UB}}\relax
\and  X. Luri         \inst{\ref{inst:UB}}\relax
\and  J. \'{A}lvarez Cid-Fuentes  \inst{\ref{inst:BSC}}\relax
\and L. Casamiquela        \inst{\ref{inst:bordeaux}}\relax
\and F. Anders         \inst{\ref{inst:UB}}\relax
\and T. Cantat-Gaudin          \inst{\ref{inst:UB}}\relax
\and M. Mongui\'o               \inst{\ref{inst:UB}}\relax
\and L. Balaguer-N\'{u}\~{n}ez         \inst{\ref{inst:UB}}\relax
\and S. Sol\`a                  \inst{\ref{inst:BSC}}\relax
\and R.M. Badia                 \inst{\ref{inst:BSC}}\relax
}

\institute{Dept. F\'isica Qu\`antica i Astrof\'isica, Institut de Ci\`encies del Cosmos (ICCUB), Universitat de Barcelona (IEEC-UB), Mart\'i i Franqu\`es 1, E08028 Barcelona, Spain\\
    \email{acastro@fqa.ub.edu}\relax \label{inst:UB}
\and
Barcelona Supercomputing Center (BSC) \relax \label{inst:BSC}
\and
Laboratoire d'Astrophysique de Bordeaux, Univ. Bordeaux, CNRS, B18N, all\'{e}e Geoffroy Saint-Hilaire, 33615 Pessac, France \relax \label{inst:bordeaux}
}

\date{Received date /
Accepted date}

\abstract{
	Open clusters are key targets for both Galaxy structure and evolution and stellar physics studies. Since \textit{Gaia} DR2 publication, the discovery of undetected clusters has proven that our samples were not complete.
}{
  	Our aim is to exploit the Big Data capabilities of machine learning to detect new open clusters in \textit{Gaia} DR2, and to complete the open cluster sample to enable further studies on the Galactic disc.
}{
  	We use a machine learning based methodology to systematically search in the Galactic disc, looking for overdensities in the astrometric space and identifying them as open clusters using photometric information. First, we use an unsupervised clustering algorithm, DBSCAN, to blindly search for these overdensities in \textit{Gaia} DR2 $(l,b,\varpi,\mu_{\alpha^*},\mu_\delta)$. After that, we use a deep learning artificial neural network trained on colour-magnitude diagrams to identify isochrone patterns in these overdensities, and to confirm them as open clusters.
}{
    We find $582$ new open clusters distributed along the Galactic disc, in the region $|b| < \ang{20}$. We can detect substructure in complex regions, and identify the tidal tails of a disrupting cluster UBC~$274$ of $\sim 3$ Gyr located at $\sim 2$ kpc.
}{
	Adapting the methodology into a Big Data environment allows us to target the search driven by physical properties of the open clusters, instead of being driven by its computational requirements. This blind search for open clusters in the Galactic disc increases in a $45\%$ the number of known open clusters.
}
\maketitle


\section{Introduction}
\label{sec:intro}

Since the publication of the second data release of the ESA mission \textit{Gaia} \citep[\textit{Gaia} DR2,][]{2016A&A...595A...1G,2018A&A...616A...1G}, which contains more than $1.3$ billion stars with precise astrometric measurements (positions, parallax and proper motions) and integrated photometry for three broad bands ($G$, $G_{BP}$ and $G_{RP}$) among other data products, the study of open clusters (OCs) has gone through a revolution with the re-definition of the OC population, in statistical terms.

Open clusters, being fundamental objects in galaxies, allow to understand our Milky Way structure and evolution. OCs are groups of stars gravitationally bound, born in the same event, therefore stars in an OC share a common position and proper motion $(l,b,\varpi,\mu_{\alpha^*},\mu_\delta)$ as well as initial chemical composition and age. The reliable estimation of ages and distances for OCs, compared to the estimation for individual stars, make them a useful tool for studying several topics in astrophysics. Young OCs allow the derivation of the initial mass function (IMF) and trace star forming regions which enable the understanding of star forming mechanisms. Intermediate to old OCs contain information about the processes occurring in the Galactic disc that disrupt these stellar structures and drive the evolution of the disc. All OCs are also indispensable to constrain stellar structure and  evolutionary models. To enable most of these studies, a complete and homogeneous census of the OC population needs to be built. 

There are many studies aimed to detect new OCs and to accurate determine of the membership. Shortly after the publication of \textit{Gaia} DR2, \citet{tristan_catalogue} was able to compute membership probabilities for $1\,229$ OCs present in catalogues previous to \textit{Gaia} DR2 \citep[where these catalogues included about $\sim 3\,000$ objects][]{dias,kharchenko}, and proved the non-existence of some of them. In parallel, \citet{acastro1} developed a machine learning (ML) methodology to search for unnoticed OCs in the \textit{Gaia} data being able to detect $23$ new OCs distributed through all sky in the TGAS data set \citep{tgas,gdr1-tgas}, and 53 new OCs in a region near the Galactic anticentre \citep{acastro2}. Since then, there have been many efforts to complete the OC census: \citet{coin_clusters} found $41$ OCs in the direction of Perseus using Gaussian mixture models, \citet{upk_clusters} could find $207$ OCs by visually inspecting proper motion diagram, \citet{2019arXiv191012600L} recently reported $2\,443$ OCs, of which $76$ were unknown and considered of high quality, by dividing the sky into small $3$-D regions and employed a friend-of-friends algorithm to search for overdensities in the $(l,b,\varpi,\mu_{\alpha^*},\mu_\delta)$ space.

So far, all these previous studies analysed either a particular region of the Galactic disc, or divided the entire Galactic disc into areas defined by the limiting number of stars that the algorithms are able to deal with due to the computational complexity and resources needed when dealing with Big Data catalogues such as \textit{Gaia}. The implementation of such methodologies into a Big Data environment, where the division of the search region of the sky into small regions depends only on the targeted structures one wants to detect, not on any computational limitation, is a key step in all sky blind searches. 

In this paper, we adapt the methodology described in \citet{acastro1} and \citet{acastro2} (CG18 and CG19 hereafter) to run in a Big Data environment. The methodology consists in the application of an unsupervised clustering algorithm, DBSCAN, to find overdensities in a five-dimensional parameter space $(l,b,\varpi,\mu_{\alpha^*},\mu_\delta)$. The confirmation of these overdensities as plausible clusters is done by recognising an isochrone pattern in the colour-magnitude diagram (CMD) of the candidates using a deep learning artificial neural network (ANN).

This paper is organised as follows. In \secref{sec:method} we discuss the methodology used, and how we adapted it to a Big Data environment. \secrefalt{sec:data} describes the data used. A review of the new OCs found is in \secref{sec:results}, as well as some general properties of the new OCs and the comparison with other OC catalogues. This section also includes some specific comments on the capabilities of the methodology. Finally, conclusions are presented in \secref{sec:conclusions}.

\section{Methodology}
\label{sec:method}

\begin{figure*}[!htb]
\centering
\includegraphics[width = 1.00\textwidth]{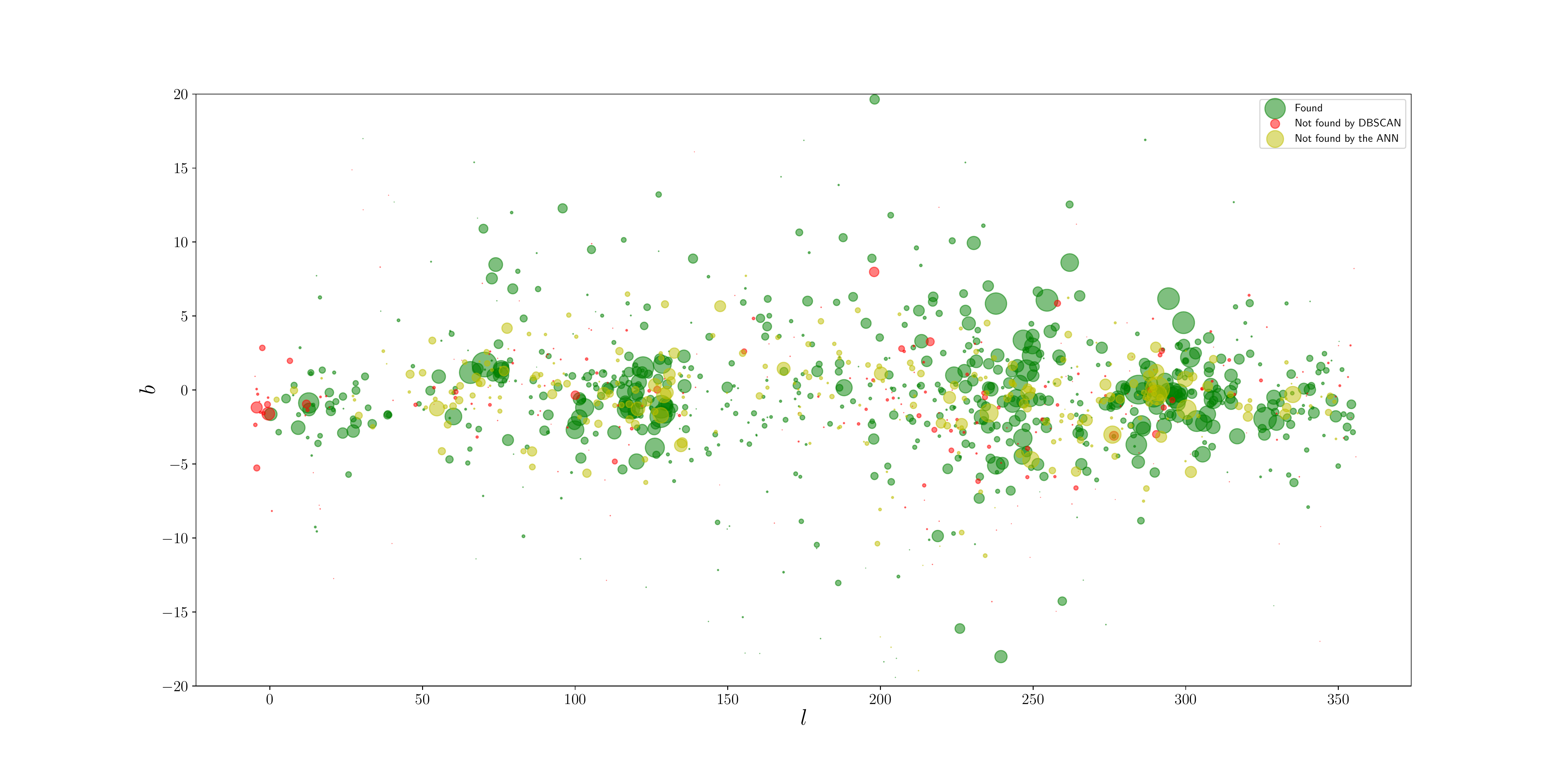}
\caption{Distribution in Galactic coordinates ($l$ \textit{vs.} $b$) of the OCs catalogued in \citet{tristan_catalogue}. Green dots represent OCs that our method recovers, red dots are OCs not found by DBSCAN and yellow dots are OCs which are found by DBSCAN but its CMD is not recognised by our ANN. The size of the dots is proportional to the star density of the cluster (see text, Eq.~\ref{eqn:vol}).}
\label{fig:check_known}
\end{figure*}

This section summarises the methodology used to systematically search for unknown OCs. The method is fully described in CG18, and it was applied to \textit{Gaia} DR2 data in CG19 to find new OCs in a region near the Galactic anticentre. 

The method consists in three main steps: preparing the data, identifying clusters with DBSCAN, and confirming them with an ANN.

In the first part, where the data is prepared, the region of search is divided into rectangles of size $L \times L$ where the five parameters $(l,b,\varpi,\mu_{\alpha^*},\mu_\delta)$ used to look for the overdensities are standardised. This division into small regions is necessary to compute an average density of the region, where the clusters located in that region represent local overdensities. Contrary to other papers, the size of these regions is defined by its homogeneity and not by the limitations of the hardware or algorithm.

Once the data is ready, the overdensities are found using a density-based clustering algorithm, DBSCAN \citep{dbscan}, which uses a statistical distance (computed as the Euclidean distance in our case) to define close by stars in 5-D as a cluster. This step has been improved with respect to CG18 and CG19 due to the larger volume of data analysed (see \secref{subsec:dislib} for details). The choice of DBSCAN is convenient because it does not require an \textit{a priori} number of clusters to be found, it is able to find arbitrarily shaped clusters and it only requires two input parameters $(\epsilon,minPts)$. The $\epsilon$ parameter is the radius of the hyper-sphere where to search for close neighbours (members of the same cluster). It is automatically computed in each $L \times L$ rectangle using the fact that the separation between stars in a cluster is smaller than between field stars (see Sect.~2.2 in CG18 for details on the computation of $\epsilon$). The parameter $minPts$ refers to the minimum number of stars within $\epsilon$ to consider them as a cluster. Once DBSCAN finds the statistical clusters in a grid defined by the $L \times L$ rectangles, the grid is shifted by $L/3$ and $2L/3$ where the algorithm is run again to account for clusters in the borders.  

The value of $minPts$ is optimised, together with $L$, using \mbox{\textit{Gaia}-like} simulated data. We used a \textit{Gaia} Universe Model Snapshot (GUMS) to simulate field stars \citep{2012A&A...543A.100R} including errors at the time of \textit{Gaia} DR2\footnote{Errors computed with the prescription given in https://github.com/agabrown/PyGaia}. OCs simulated using the \textit{Gaia} Object Generator (GOG) \citep{gog} were added to the GUMS simulation as the objects to be found by DBSCAN. A pair of $(L,minPts)$ is considered to be optimal if a balance is reached in terms of low contamination and high efficiency.

For true data, the whole process is run over the several $(L,minPts)$ optimal parameters to assess for the reliability of the clusters found. The more times a statistical cluster has been found within the explored $(L,minPts)$ pairs, the more likely to be a real OC. The values of $(L,minPts)$ used are $35$ combinations of $L \in [\ang{9},\ang{15}] \text{ and } minPts \in [8,16]$.

As a last step, overdensities found with DBSCAN are classified into real OCs or just statistical clusters using an ANN \citep{ann}, trained to recognise the characteristic isochrone pattern of OCs in the CMD. This step has also been improved with respect to CG18 and CG19, resulting in a more robust classification with the use of deep learning (see \secref{subsec:deeplearning}).  

\subsection{Distributed computation of DBSCAN}
\label{subsec:dislib}

So far, the method had been applied to small volume data sets (\textit{i.e.} to TGAS in CG18, and to a region in the Galactic anticentre up to magnitude $G = 17$ in CG19) for design and validation purposes. Both previous studies used the DBSCAN implementation from scikit-learn \citep{sklearn}, an easy-to-use API that provides machine learning algorithms for Python. However, the higher stellar density to be analysed in other regions of the disc, towards the Galactic centre for instance, requires a machine learning library able to be deployed in a distributed environment and to handle larger volumes of data.

In this paper, we have used PyCOMPSs \citep{pycompss} to find overdensities in the whole Galactic disc $(\ang{0} \leq l \leq \ang{360}$ and $\ang{-20} \leq b \leq \ang{20})$ down to magnitude $G = 17$. PyCOMPSs is a task-based programming model that automatically manages the distribution of the computation depending on the available resources. Using PyCOMPSs, we build an application that uses scikit-learn's DBSCAN to different regions of the Galactic disc in parallel. This speeds up the computation time, and allows us to process a volume of data that does not fit in the memory of a single machine.

The algorithm is deployed on the MareNostrum 4 supercomputer\footnote{https://www.bsc.es/marenostrum} installed at the Barcelona Supercomputing Center (BSC). The nodes used for the computation of DBSCAN have $96\,GB$ of memory and $48$ cores per node. For performance comparison purposes, we run DBSCAN with the same configuration that we used in CG18 on the TGAS data set. In that case, in CG18, the computation of DBSCAN for all the optimal parameters took $18$ hours in a sequential execution in a single machine, whereas using PyCOMPSs the whole computation takes $\sim 1.4$ hours in $1$ node ($48$ cores) and less than $18$ minutes in $4$ nodes ($192$ cores) \citep[see Sect.~5 from][for a detailed comparison]{dislib}.

For this case, the analysis of the whole Galactic disc (defined as $\ang{-20} \leq b \leq \ang{20}$) up to magnitude $G = 17$ using DBSCAN on $4$ nodes ($192$ cores) takes an average of $8.27$ hours per pair of parameters, ranging from $5.67$ to $11.17$ hours depending on the pairs of $(L,minPts)$.  

\subsection{OCs validation with Deep Learning}
\label{subsec:deeplearning}

The application of DBSCAN over a large volume of data, with several optimal pairs of parameters $(L,minPts)$ picks up a large number of statistical overdensities which correspond to real OCs, also including overdensities only in statistical terms. To automatically decide if a given statistical cluster is a real OC we have trained an ANN to recognise the isochrone patterns that stars in OCs follow in a CMD. For both CG18 and CG19 we used a simple multi-layer perceptron with one hidden layer to make the classification. In this paper, due to the large amount of statistical clusters found, a more complex model is needed for robust classification. We have designed a (deep-)ANN, including several convolutional layers, to perform the classification.

The (deep-)ANN is implemented in PyTorch\footnote{https://pytorch.org/} \citep{pytorch}, a popular and powerful deep learning library. It takes as input a two dimensional histogram in $G_{BP}-G_{RP}$ \textit{vs} $G$, \textit{i.e.} a CMD, and is trained to decide whether it belongs to a real OC or not. The network is built in two blocks; a first block consisting in a set of convolutional layers which are able to learn the features and geometry of the isochrone pattern in the CMD, and a second block with two fully connected layers where the classification of the learned features is performed. After each layer, a ReLU activation function ($f(x) = \text{max}(0,x)$) is added, which has proved to give better results than other activation functions \citep{relu}.

\subsubsection{Building the training set}
\label{subsubsec:training}

One of the caveats of deep learning is that it requires a large amount of training samples to learn the possible configurations of the feature space. The CMDs of the $\sim1\,500$ confirmed OCs are not sufficient to train the network. Moreover, some of these OCs have not enough stars ($minPts$ at least) with magnitudes $G \leq 17$ or the isochrone is very dispersed, so we had to remove these clusters from the training set. To enlarge the training set we used data augmentation techniques (see description in Sect.~2.3.2 in CG18) on the real known OCs. In addition to the known OCs, we use simulated isochrones from the \mbox{PARSEC} code \citep{2012MNRAS.427..127B}. To build the set of isochrones, we assume solar metallicity $(Z \simeq 0.0152)$ and ages ranging from log$(age) = 6.6$ dex to log$(age) = 10.3$ dex in steps of $0.1$ dex. For each age, the isochrone is filled with a population of a total mass of $10^4$ M$_\odot$ following the IMF described in \citet{2001MNRAS.322..231K}. Then, we select different sub samples of the whole population to create the simulated OCs, and we locate them at different distances (ranging from $0.4$ to $4$ kpc) to better represent the parameter space. For each sub sample, the CMD is built in the $G_{BP}-G_{RP}$ \textit{vs.} $G$ space using the photometric  pass bands described in \citet{2018A&A...619A.180M}. Finally, in order to mimic \textit{Gaia} DR2 results, we add photometric errors \citep{2018A&A...616A...4E} using an analytical prescription provided by Carrasco et al. (private communication) and a fraction of binaries. On the negative identification side, CMDs from random (field) stars located at different fields in the whole studied area are used.

Each CMD is converted to a $2$-D histogram, and as a preprocessing step, we normalise the data (each pixel of the histogram is limited between $0$ and $1$) before feeding the whole $2$-D histogram to the network. To reach better classification performance, a logarithmic normalisation was done in order to highlight the lower density regions so the network takes into account the contamination from field stars when performing the classification.

\subsubsection{Performance of the classification}
\label{subsubsec:performance}

The performance of the classification is assessed in two steps. On the one hand, the whole training set is split into training and test with $80\% - 20\%$ of the whole set, respectively. This is useful when designing the network architecture because the true classification of each sample is known. The final architecture is chosen to be the one that minimises the test loss.

On the other hand, the model is applied to the anticentre area as in CG19, where we found $53$ new OCs from $491$ candidates. We do not know the true classification of each of those $491$ samples, so the final parameters of the ANN here are tuned to keep a $80\%$ (at least) of the OCs confirmed in that region, minimising the manually discarded statistical clusters. When applying the final model to classify all the statistical clusters found in the Galactic disc, we can recover this $80\%$ requirement (in terms of known OCs recovered) showing that the results will be equivalent in both sets.

\begin{figure*}[!htb]
\begin{subfigure}{1.00\textwidth}
\includegraphics[width = 1.\textwidth]{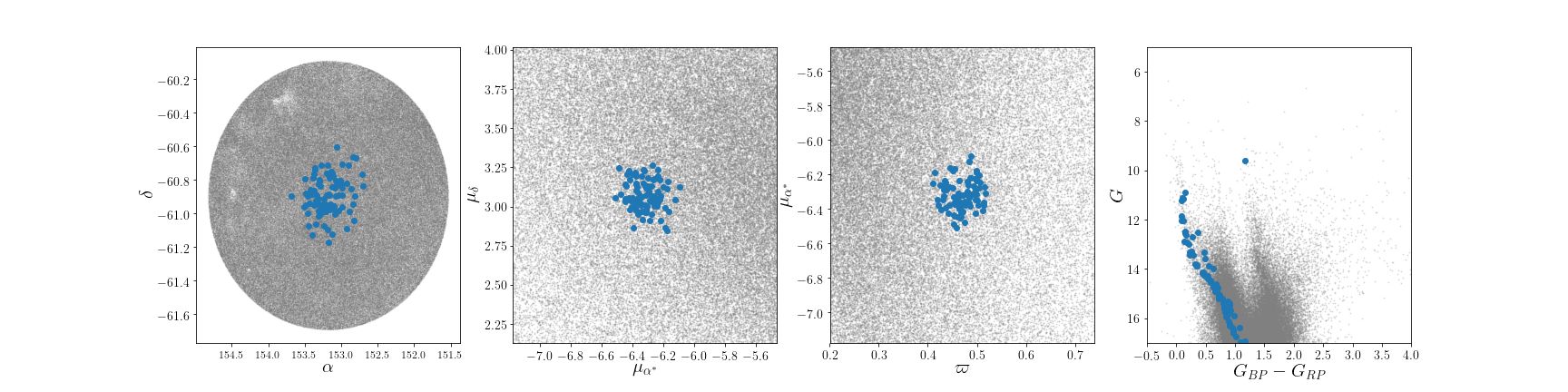}
\end{subfigure}
\begin{subfigure}{1.00\textwidth}
\includegraphics[width = 1.\textwidth]{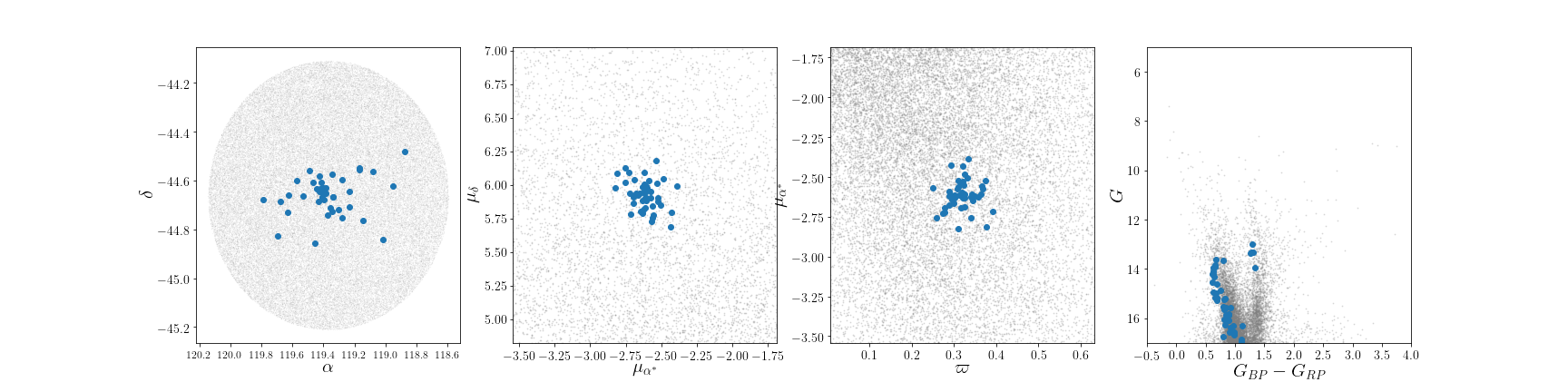}
\end{subfigure}
\begin{subfigure}{1.00\textwidth}
\includegraphics[width = 1.\textwidth]{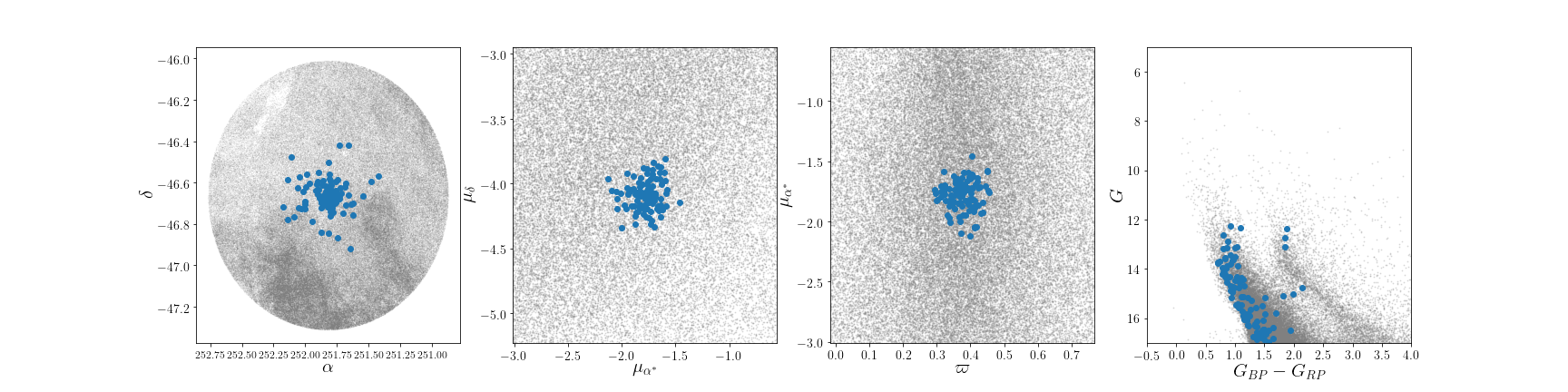}
\end{subfigure}
\caption{Examples of class A (top row), class B (middle row) and class C (bottom row) clusters. The columns represent, from left to right, a distribution of the member stars (in blue) and field stars (in grey) for: i) position in $(\alpha,\delta$), ii) proper motions in $(\mu_{\alpha^*},\mu_\delta)$, iii) distribution in $(\varpi,\mu_{\alpha^*})$ and iv) a CMD in $G$ \textit{vs.} $G_{BP}-G_{RP}$. Figures correspond to OCs UBC~$257$, UBC~$478$ and UBC~$669$, respectively. Classes A, B and C correspond to different levels of reliability (see \secref{subsec:new}).}
\label{fig:examples}
\end{figure*}

\section{Data}
\label{sec:data}

The data used to perform the blind search for OCs is the \textit{Gaia} DR2 \citep{2018A&A...616A...1G}. In its second data release, \textit{Gaia} provides precise astrometry and kinematics $(l,b,\varpi,\mu_{\alpha^*},\mu_\delta)$ in addition to excellent photometry in three broad bands $(G,G_{BP},G_{RP})$. The search is focused in the Galactic disc, defined as $\ang{0} \leq l \leq \ang{360}$ and $\ang{-20} \leq b \leq \ang{20}$, because the expectation of finding OCs in that region is maximum; \textit{i.e.} $99\%$ of the known OCs catalogued in \citet{tristan_catalogue} are in $|b| < \ang{20}$, similarly for \citet{dias} and \citet{kharchenko} with $96\%$ and $94\%$ of the total reported objects in $|b| < \ang{20}$, respectively.

The data set is also limited in magnitude, up to $G = 17$, where the median astrometric uncertainties are $0.094$ mas for the parallax, $0.158$ and $0.137$ mas yr$^{-1}$ for $\mu_{\alpha^*}$ and $\mu_\delta$, respectively \citep{2018A&A...616A...2L}. On the photometric side, up to magnitude $G = 17$ the uncertainties are at the level of $\sim 0.001$ mag for $G$, $\sim 0.006$ mag for $G_{BP}$ and $\sim 0.01$ mag for $G_{RP}$ \citep{2018A&A...616A...4E}. We consider these uncertainty levels to be good limits to have proper results with our method. This results in a sample containing $122\,727\,809$ stars.

\section{Results}
\label{sec:results}

\begin{table*}
\caption{Few examples of the proposed OCs ordered by increasing $l$. The parameters shown are the mean (and standard deviation) for the ($N$) members found also including the apparent angular size ($\theta$) and estimated distance ($d$) with one sigma confidence interval. Radial velocity is included when available and it is computed with $N_{V_{rad}}$ members. The name follows the numeration of CG19. The full list can be found online at the CDS.}
\label{tab:new_oc}
\centering
\resizebox{\textwidth}{!}{
\renewcommand{\arraystretch}{1.15}
\begin{threeparttable}[b]
\begin{tabular}{cccccccccccc}
\hline
\hline
\multicolumn{1}{c}{Name} &\multicolumn{1}{c}{\begin{tabular}[c]{@{}c@{}}$\alpha$\\ 
$[\Unit{deg}]$\end{tabular}} &\multicolumn{1}{c}{\begin{tabular}[c]{@{}c@{}}$\delta$\\ 
$[\Unit{deg}]$\end{tabular}} &\multicolumn{1}{c}{\begin{tabular}[c]{@{}c@{}}$l$\\ 
$[\Unit{deg}]$\end{tabular}} &\multicolumn{1}{c}{\begin{tabular}[c]{@{}c@{}}$b$\\ 
$[\Unit{deg}]$\end{tabular}} &\multicolumn{1}{c}{\begin{tabular}[c]{@{}c@{}}$\theta$\\ 
$[\Unit{deg}]$\end{tabular}} &\multicolumn{1}{c}{\begin{tabular}[c]{@{}c@{}}$\parallax$\\ 
$[\Unit{mas}]$\end{tabular}} &\multicolumn{1}{c}{\begin{tabular}[c]{@{}c@{}}$d$\\ 
$[\Unit{kpc}]$\end{tabular}} &\multicolumn{1}{c}{\begin{tabular}[c]{@{}c@{}}$\mu_{\alpha^*}$\\ 
$[\Unit{mas\cdot yr}^{-1}]$\end{tabular}} & \multicolumn{1}{c}{\begin{tabular}[c]{@{}c@{}}$\mu_{\delta}$\\ 
$[\Unit{mas\cdot yr}^{-1}]$\end{tabular}}  & \multicolumn{1}{c}{\begin{tabular}[c]{@{}c@{}}$V_{\rm rad}$\\ 
$[\Unit{km\cdot s}^{-1}]$\end{tabular}}  & \multicolumn{1}{c}{$N$ ($N_{V_{\rm rad}}$)} \\ 
\hline 
\multicolumn{10}{c}{Class A}\\ 
\hline 
UBC~$91$&$267.42$$(0.07)$&$-28.76$$(0.07)$&$0.61$$(0.07)$&$-0.67$$(0.06)$&$0.09$&$0.42$$(0.03)$&$2.37^{+0.18}_{-0.16}$&$-0.59$$(0.09)$&$-1.12$$(0.11)$&$-$$(-)$&$83$$(0)$\\ 
UBC~$92$&$269.88$$(0.07)$&$-26.65$$(0.06)$&$3.53$$(0.07)$&$-1.49$$(0.06)$&$0.09$&$0.38$$(0.04)$&$2.66^{+0.31}_{-0.25}$&$2.13$$(0.09)$&$0.41$$(0.09)$&$-10.79$$(2.85)$&$105$$(2)$\\ 
UBC~$93$&$268.57$$(0.05)$&$-25.39$$(0.05)$&$4.03$$(0.04)$&$0.17$$(0.05)$&$0.07$&$0.34$$(0.03)$&$2.95^{+0.25}_{-0.22}$&$-0.93$$(0.11)$&$-1.88$$(0.09)$&$-$$(-)$&$52$$(0)$\\ 
UBC~$94$&$269.63$$(0.09)$&$-24.64$$(0.1)$&$5.17$$(0.1)$&$-0.29$$(0.08)$&$0.13$&$0.75$$(0.01)$&$1.34^{+0.03}_{-0.02}$&$-1.66$$(0.07)$&$-4.45$$(0.06)$&$-$$(-)$&$41$$(0)$\\ 
UBC~$95$&$268.25$$(0.06)$&$-22.17$$(0.09)$&$6.66$$(0.09)$&$2.06$$(0.07)$&$0.11$&$0.49$$(0.03)$&$2.03^{+0.12}_{-0.1}$&$-0.15$$(0.13)$&$-1.28$$(0.11)$&$-16.16$$(-)$&$84$$(1)$\\ 
UBC~$96$&$273.76$$(0.09)$&$-16.33$$(0.1)$&$14.31$$(0.11)$&$0.39$$(0.08)$&$0.14$&$0.62$$(0.02)$&$1.62^{+0.07}_{-0.06}$&$0.64$$(0.11)$&$0.93$$(0.08)$&$-$$(-)$&$41$$(0)$\\ 
UBC~$97$&$274.78$$(0.1)$&$-15.73$$(0.08)$&$15.3$$(0.1)$&$-0.18$$(0.08)$&$0.12$&$0.73$$(0.02)$&$1.36^{+0.03}_{-0.03}$&$-0.87$$(0.08)$&$-1.15$$(0.08)$&$-$$(-)$&$33$$(0)$\\ 
UBC~$98$\tnote{a}&$288.83$$(0.15)$&$-22.14$$(0.14)$&$15.38$$(0.13)$&$-14.93$$(0.16)$&$0.2$&$1.53$$(0.03)$&$0.65^{+0.01}_{-0.01}$&$0.56$$(0.11)$&$-6.66$$(0.17)$&$-$$(-)$&$23$$(0)$\\ 
UBC~$99$\tnote{a}&$282.02$$(0.09)$&$-18.3$$(0.09)$&$16.18$$(0.08)$&$-7.52$$(0.09)$&$0.13$&$1.06$$(0.03)$&$0.94^{+0.03}_{-0.03}$&$-1.16$$(0.1)$&$-4.1$$(0.13)$&$-$$(-)$&$52$$(0)$\\ 
UBC~$100$&$281.26$$(0.07)$&$-11.12$$(0.1)$&$22.3$$(0.1)$&$-3.65$$(0.07)$&$0.12$&$0.7$$(0.01)$&$1.43^{+0.03}_{-0.03}$&$-1.1$$(0.08)$&$-3.33$$(0.09)$&$-$$(-)$&$25$$(0)$\\ 
UBC~$101$&$279.5$$(0.09)$&$-7.14$$(0.07)$&$25.05$$(0.08)$&$-0.28$$(0.08)$&$0.11$&$0.42$$(0.02)$&$2.41^{+0.15}_{-0.13}$&$-0.31$$(0.09)$&$-3.03$$(0.08)$&$15.89$$(-)$&$54$$(1)$\\ 
UBC~$102$&$280.61$$(0.08)$&$-6.89$$(0.09)$&$25.77$$(0.09)$&$-1.15$$(0.08)$&$0.12$&$0.52$$(0.02)$&$1.94^{+0.08}_{-0.08}$&$-1.04$$(0.09)$&$-2.51$$(0.11)$&$9.97$$(-)$&$42$$(1)$\\ 
UBC~$103$&$280.63$$(0.05)$&$-6.6$$(0.08)$&$26.04$$(0.07)$&$-1.04$$(0.06)$&$0.09$&$0.28$$(0.03)$&$3.54^{+0.35}_{-0.29}$&$-0.4$$(0.09)$&$-2.27$$(0.09)$&$-3.99$$(-)$&$97$$(1)$\\ 
UBC~$104$&$280.69$$(0.05)$&$-6.26$$(0.07)$&$26.37$$(0.06)$&$-0.93$$(0.06)$&$0.08$&$0.29$$(0.03)$&$3.45^{+0.44}_{-0.35}$&$0.49$$(0.09)$&$-0.8$$(0.09)$&$-1.25$$(2.17)$&$61$$(2)$\\ 
UBC~$105$&$280.33$$(0.09)$&$-5.43$$(0.08)$&$26.94$$(0.08)$&$-0.23$$(0.08)$&$0.12$&$0.47$$(0.03)$&$2.14^{+0.12}_{-0.11}$&$0.46$$(0.11)$&$-0.99$$(0.09)$&$-$$(-)$&$75$$(0)$\\ 
\multicolumn{10}{c}{$\vdots$}\\ 
\hline 
\multicolumn{10}{c}{Class B}\\ 
\hline 
UBC~$336$&$267.98$$(0.03)$&$-27.83$$(0.03)$&$1.66$$(0.03)$&$-0.62$$(0.03)$&$0.04$&$0.31$$(0.02)$&$3.2^{+0.18}_{-0.16}$&$0.75$$(0.08)$&$0.14$$(0.07)$&$-25.48$$(-)$&$22$$(1)$\\ 
UBC~$337$&$271.72$$(0.08)$&$-24.65$$(0.08)$&$6.09$$(0.07)$&$-1.94$$(0.08)$&$0.11$&$0.57$$(0.02)$&$1.77^{+0.06}_{-0.06}$&$0.47$$(0.08)$&$-0.72$$(0.07)$&$-$$(-)$&$40$$(0)$\\ 
UBC~$338$&$271.53$$(0.07)$&$-24.23$$(0.08)$&$6.37$$(0.08)$&$-1.59$$(0.06)$&$0.1$&$0.6$$(0.02)$&$1.66^{+0.06}_{-0.06}$&$0.01$$(0.08)$&$-1.77$$(0.09)$&$-15.86$$(-)$&$38$$(1)$\\ 
UBC~$339$&$271.31$$(0.04)$&$-23.31$$(0.05)$&$7.08$$(0.05)$&$-0.96$$(0.04)$&$0.06$&$0.39$$(0.02)$&$2.59^{+0.12}_{-0.11}$&$0.57$$(0.07)$&$-0.59$$(0.08)$&$-$$(-)$&$19$$(0)$\\ 
UBC~$340$&$270.77$$(0.09)$&$-22.66$$(0.07)$&$7.4$$(0.06)$&$-0.21$$(0.09)$&$0.11$&$0.7$$(0.02)$&$1.42^{+0.03}_{-0.03}$&$0.72$$(0.07)$&$-2.57$$(0.08)$&$-$$(-)$&$27$$(0)$\\ 
UBC~$341$&$276.45$$(0.1)$&$-17.06$$(0.09)$&$14.87$$(0.1)$&$-2.23$$(0.08)$&$0.13$&$0.48$$(0.03)$&$2.1^{+0.13}_{-0.11}$&$-0.21$$(0.12)$&$-1.49$$(0.1)$&$-3.75$$(-)$&$94$$(1)$\\ 
UBC~$342$&$273.91$$(0.17)$&$-14.92$$(0.17)$&$15.61$$(0.13)$&$0.94$$(0.2)$&$0.24$&$0.6$$(0.03)$&$1.66^{+0.09}_{-0.08}$&$-0.17$$(0.11)$&$-1.04$$(0.14)$&$-$$(-)$&$66$$(0)$\\ 
\multicolumn{10}{c}{$\vdots$}\\ 
\hline 
\multicolumn{10}{c}{Class C}\\ 
\hline 
UBC~$572$&$280.42$$(0.07)$&$-21.95$$(0.06)$&$12.2$$(0.06)$&$-7.78$$(0.07)$&$0.09$&$0.65$$(0.02)$&$1.54^{+0.05}_{-0.05}$&$0.98$$(0.1)$&$-0.63$$(0.11)$&$-33.02$$(7.31)$&$23$$(2)$\\ 
UBC~$573$&$275.01$$(0.07)$&$-9.44$$(0.09)$&$20.95$$(0.09)$&$2.58$$(0.07)$&$0.11$&$0.53$$(0.02)$&$1.88^{+0.09}_{-0.08}$&$-0.18$$(0.1)$&$-4.48$$(0.1)$&$-$$(-)$&$17$$(0)$\\ 
UBC~$574$\tnote{a}&$282.32$$(0.08)$&$-4.36$$(0.09)$&$28.8$$(0.09)$&$-1.51$$(0.08)$&$0.12$&$0.58$$(0.0)$&$1.73^{+0.01}_{-0.01}$&$1.06$$(0.02)$&$0.21$$(0.04)$&$-10.15$$(-)$&$9$$(1)$\\ 
UBC~$575$&$291.01$$(0.08)$&$-5.13$$(0.11)$&$32.05$$(0.1)$&$-9.58$$(0.09)$&$0.13$&$0.91$$(0.02)$&$1.09^{+0.02}_{-0.02}$&$-0.3$$(0.07)$&$-5.18$$(0.08)$&$-$$(-)$&$9$$(0)$\\ 
UBC~$576$&$284.68$$(0.04)$&$0.42$$(0.06)$&$34.13$$(0.06)$&$-1.43$$(0.04)$&$0.07$&$0.74$$(0.02)$&$1.34^{+0.03}_{-0.03}$&$-0.74$$(0.08)$&$-3.56$$(0.09)$&$-$$(-)$&$17$$(0)$\\ 
UBC~$577$&$282.17$$(0.05)$&$22.12$$(0.09)$&$52.54$$(0.09)$&$10.47$$(0.05)$&$0.1$&$1.0$$(0.02)$&$1.0^{+0.02}_{-0.02}$&$-1.04$$(0.11)$&$3.07$$(0.07)$&$-0.59$$(16.75)$&$9$$(4)$\\ 
\multicolumn{10}{c}{$\vdots$}\\ 
\hline 
\end{tabular}\begin{tablenotes}
\item[a] coincidence with \citet{upk_clusters} or \citet{2019arXiv191012600L}, see \secref{subsubsec:sim}
\item[b] tentative identification with \citet{kharchenko}, see \secref{subsubsec:dias+kharchenko} 
\end{tablenotes}
\end{threeparttable}
}
\end{table*}

The described methodology is applied to the whole Galactic disc. This results in a list of $2\,213$ possible OC candidates, including the already known OCs and newly discovered ones.

\subsection{Comparison with existing catalogues}
\label{subsec:comparison}

To report only newly discovered OCs, we cross-match our list of detections with other catalogues to see which groups are already known.

\subsubsection{Cantat-Gaudin et al. (2018)}
\label{subsubsec:checkTristan}

We consider a candidate to be matched with one OC in the \citet{tristan_catalogue} catalogue if their mean parameters are compatible within $2\sigma_i$, where $\sigma_i$ is the standard deviation computed from the members of each OC in the five dimensional astrometric space, $i = \{l,b,\varpi,\mu_{\alpha^*},\mu_\delta\}$. From our $2\,213$ OC candidates, $688$ are listed in \citet{tristan_catalogue} with our matching criteria. This represents a $\sim 81\%$ of the OCs reported in \citet{tristan_catalogue} used in the training set for the ANN, where we removed OCs either with few members up to $G = 17$ or with not well defined empirical isochrones in the CMDs that would confuse the ANN for the classification.

Our strategy to compute the DBSCAN parameters $(L,minPts)$ relies in the higher star density of a cluster, compared to field stars. Therefore, our detection is limited to the most compact objects in the field of search ($L \times L$). This is seen in \figref{fig:check_known}, where a distribution of \mbox{$l$ \textit{vs.} $b$} of the catalogued OCs is shown. The OCs found by our method are plotted in green, whereas those not found are plotted either in red (if it is not found by DBSCAN) or in yellow (if its sequence in the CMD is not well defined, and so not recognised by our ANN). The size of the dots is proportional to the density of the cluster in the five-dimensional astrometric space, computed as the $68\%$ of the total number of stars of the cluster divided by the volume of a 5-D hyper-sphere:

\begin{equation}
\label{eqn:vol}
V_5 = \frac{\pi^{\frac{5}{2}}}{\Gamma (\frac{5}{2} + 1)} r^5,
\end{equation}
where $r = (\sigma_l^2 + \sigma_b^2 + \sigma_\varpi^2 + \sigma_{\mu_{\alpha^*}}^2 + \sigma_{\mu_\delta}^2)^{\frac{1}{2}}$ for each cluster. Found OCs are mostly high density groups, whilst those not found are either low density objects (which are near a higher density object), or its sequence in the CMD is not recognised as an isochrone by our ANN.

\subsubsection{Castro-Ginard et al. (2018) and (2019), and Cantat-Gaudin et al. (2019)}
\label{subsubsec:alfred}

The method discussed in this paper was presented in CG18, where a blind search was performed over the TGAS data \citep{gdr1-tgas}. The $23$ OCs found in CG18, mainly closer than $1$ kpc (due to the bright limiting magnitude), are not likely to be found with \textit{Gaia} DR2 due to the very different star density of the data set and the parameters $(L,minPts)$ used in the search. However, we can find UBC~$3$, UBC~$6$, UBC~$8$, UBC~$9$ and UBC~$27$.

CG19 and \citet{coin_clusters} applied different methodologies to an area covering the Galactic anticentre. They found $53$ and $41$ previously unknown OCs, respectively, with $21$ OCs in common. They found that the techniques are complementary, with none of the explored methods able to detect all the objects.

These studies analysed a very particular region of the disc, where the star density is low compared to any other disc region. In the present work, we are able to find $42$ out of the $53$ (\textit{i.e} $80\%$) OCs found in CG19 using the same methodology. The reason for not finding the $11$ OCs left is because the parameters $(L,minPts)$ used in the DBSCAN search (in the case of CG19) were optimised for that region of low stellar density. When optimising these parameters for a blind search in the whole Galactic disc, one has to account for regions with very different stellar densities. The optimal parameters chosen here are those that show the best performance in general terms, reaching a balance between low and high density regions. For the case of \citet{coin_clusters}, we could only find $24$ out of the $41$ reported OCs for similar reasons.

\subsubsection{Dias et al. (2002) and Kharchenko et al. (2013)}
\label{subsubsec:dias+kharchenko}

These catalogues contain $\sim 3\,000$ OCs each compiled from heterogeneous data sources, which difficults the cross-match with our candidates. A candidate is considered to be tentatively matched with one object in those catalogues if its centres lie within a circle of radius $\ang{0.5}$. If two objects are tentatively matched by this positional criterium, we check if the mean values in $(\mu_{\alpha^*},\mu_\delta)$ are compatible by performing a Welch t-test \citep{10.1093/biomet/34.1-2.28}, with a threshold p-value $= 0.05$ to reject the null hypothesis (to reject that they are compatible). To perform the Welch t-test, we take the \citet{kharchenko} most probable members for the cluster central part as the number of members for each OC in \citet{kharchenko}. These catalogues do not report the mean parallax for each OC but an estimation of the distance instead, with no uncertaity associated. Therefore no comparison is done in this dimension.

Most of the coincidences with these catalogues have been already taken into account by the cross-match of our candidates with \citet{tristan_catalogue}. However, we find $5$ OCs that are compatible with the position and proper motion (with p-value $> 0.05$) criteria described above. Those objects are flagged in our \tabref{tab:new_oc}.

With our methodology we are also able to identify objects related with known star forming regions. Some of them are listed in the aforementioned catalogues. We find objects which are related with $\sigma$-Ori, Collinder~$228$, Bochum~$10$, NGC~$1980$, NGC~$1981$, NGC~$6514$, NGC~$6530$ and NGC~$6604$ \citep{2008hsf1.book.....R,2008hsf2.book.....R}. While $\sigma$-Ori is listed as a possible stellar association in \citet{dias} it is considered a moving group in \citet{kharchenko}. Collinder~$228$ has variable extinction according to \citet{dias} and has nebulosity according to \citet{kharchenko}. Bochum~$10$ and NGC~$6604$ are normal clusters in both catalogues. NGC~$1980$ and $1918$ are considered a normal OC and an embedded OC in a possible OB association, respectively, in \citet{dias} while they are considered as nebulosities in \citet{kharchenko}. Finally, NGC~$6514$ and NGC~$6530$ are listed as normal OCs in \citet{dias} and as nebulosities in \citet{kharchenko}.

\subsubsection{Bica et al. (2019)}
\label{subsubsec:bica}

\citet{2019AJ....157...12B} compiled a catalogue with $10\,978$ stellar clusters, associations and candidates reported previous to \textit{Gaia} DR2, by combining together catalogues from different studies on different surveys (Digital Sky Survey, 2MASS, WISE, VVV, Spitzer and Herschel). Among the groups listed by \citet{2019AJ....157...12B}, the OCs amount to $\sim 3\,000$. Others are $\sim 300$ globular clusters, $\sim 5\,000$ embedded clusters which are hardly seen by \textit{Gaia} and $\sim 1200$ asterisms. The coincidences among OCs are already discussed in the previous subsections. We find $45$ additional coincidences with their catalogue. These matches correspond to globular clusters (GC), which \citet{2019AJ....157...12B} include, and where not taken into account in the previous cross-matches.

The detection of GCs by our methodology is a good diagnostic test. On the one hand, DBSCAN is able to detect these GCs repeatedly among all the DBSCAN runs (for all optimal $L,minPts$ parameters). For instance $\omega$-Cen, the most massive GC known with $4\times10^6$M$\odot$, is the cluster found more times by our algorithm. On the other hand, the ANN was trained with CMDs from real OCs and from simulated stellar populations at different ages. Since OCs are mostly young objects, the contribution to the recognition of such an old isochrone ($>10$ Gyr) comes from the simulated data (with the appropriate error model). Therefore, the use of simulated CMDs not only contribute in increasing the training set, but it allows the ANN to recognise cases in the real data that were trained using simulations. 

\subsubsection{Sim et al. (2019) and Liu et al. (2019)}
\label{subsubsec:sim}

Recently, \citet{upk_clusters} found $207$ new OCs, located within $1$ kpc, by visually inspecting \textit{Gaia} DR2 proper motion diagrams searching for overdensities. The criteria to consider one of these objects to be matched with one of our candidates is similar to the previous section. We consider a tentative identification if the centres of both objects lie within a circle of radius $\ang{0.5}$ and then we compare the rest of the astrometric parameters.

Firstly, we find that one of these objects, UPK~$19$, corresponds to UBC~$32$, already reported by CG18. In this case, UPK~$19$ and UBC~$32$ are separated by $\ang{0.18}$ in the sky and the rest of their mean astrometric parameters differ by $(2\sigma_{\varpi}, 0.14\sigma_{\mu_{\alpha^*}}, 0.15\sigma_{\mu_\delta})$. 

Secondly, $8$ of our OC candidates are identified with one UPK object. All the identifications are compatible within $1\sigma$ in proper motions. The mean parallaxes are compatible within $1.91\sigma$ (at most), this larger discrepancy is because \citet{upk_clusters} do not report mean parallaxes but the estimated distance instead, and the transformation from parallax to distance may lead to big differences. However, we consider these objects as matched.

Similarly, \citet{2019arXiv191012600L} identified $2\,443$ star clusters in the Galactic disc using a clustering algorithm in the 5-D astrometric space $(l,b,\varpi,\mu_{\alpha^*},\mu_\delta)$. Most of these star clusters were previously reported. $76$ of their high confidence candidates are reported as new objects. Among these $76$, we find $4$ coincidences with CG18 and CG19. These are the cases for their clusters with \textit{id}s: $1973, 2143, 2230, 2385$ which are identified with UBC~$74$, UBC~$72$, UBC~$56$, UBC~$7$ (from CG18 and CG19), respectively. All the identifications are closer than $\ang{0.5}$ in and within $2\sigma$ in $(\varpi,\mu_{\alpha^*},\mu_\delta)$. From our list of new OC candidates, we find $45$ cases that are compatible with one of the $76$ from \citet{2019arXiv191012600L}, with the same matching criteria.

\begin{figure*}[!htb]
\includegraphics[width = 1.\textwidth]{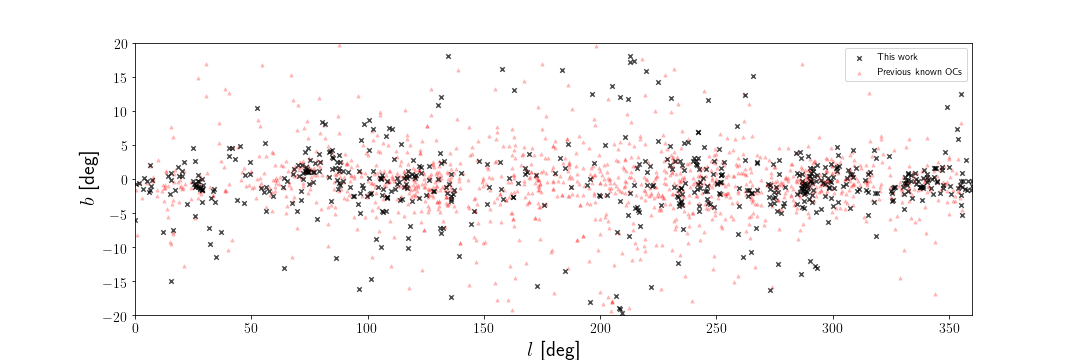}
\caption{Distribution of the OC census in $l$ \textit{vs.} $b$. Green crosses represent new OCs while blue triangles represent OCs in CG18, CG19, \citet{tristan_catalogue} and \citet{coin_clusters}.}
\label{fig:lbplane}
\end{figure*}

\subsection{Newly found OCs}
\label{subsec:new}

We select as new OCs those candidates which are found more than $3$ times among all the runs we have applied the method (each time with a different set of optimal parameters $(L,minPts)$, see \secref{sec:method}). This results in a list of $676$ tentative new structures.

These structures are further divided into three categories: new OCs of class A, class B and class C; plus other stellar structures that were discarded. We classify the new OCs into these categories by visually inspecting the CMD of the candidates, and the distribution of their member stars in the astrometric space (\figref{fig:examples}), including radial velocity when available. \tabref{tab:new_oc}\footnote{Full version, with the $582$ OCs, available online at the CDS.} lists the mean parameters of the candidates proposed as OCs $(\alpha,\delta,l,b,\varpi,\mu_{\alpha^*},\mu_\delta,V_{rad})$ as well as the apparent angular size computed as $\theta = \sqrt{\sigma_l^2+\sigma_b^2}$. An estimation of the distance by the inversion the mean parallax is also included, with (asymmetric) confidence intervals. A list with the members for each OC, as computed by DBSCAN, is available in Table~2\footnote{Table~2 is only available online at the CDS.}.

These categories count with $245$ OCs in class A, $236$ in class B and $101$ in class C. \tabref{tab:mean_classes} shows the mean $(\theta,\varpi,\sigma_{\mu_{\alpha^*}},\sigma_{\mu_\delta},N,N_\text{found})$ for each class. \figrefalt{fig:examples} shows one OC of each category. Class A clusters, typically show a high concentration of the member stars in all the five astrometric parameters $(l,b,\varpi,\mu_{\alpha^*},\mu_\delta)$, and a clean isochrone in a CMD. Clusters in class B show a more sparse distribution in the five astrometric parameters, and many include a low number of contaminant (field) stars which can be seen more clearly in the CMD. While clusters in class C are typically poorly populated and show an isochrone that could have a higher degree of contaminant stars. From the OCs classified as class A, $115$ $(47\%)$ have stars evolved beyond the main sequence, this represents the oldest population of this class.

\begin{table}
\renewcommand\thetable{3}
\caption{Mean parameters for each of the OC classes. The showed parameters are: angular size, parallax, proper motions, number of members and number of times found within all runs of the method.}
\label{tab:mean_classes}
\centering
\begin{tabular}{ccccccc}
\hline
\hline
  & $\theta$   &  $\varpi$ & $\sigma_{\mu_{\alpha^*}}$  & $\sigma_{\mu_\delta}$  & $N$  & $N_\text{found}$ \\
\hline
Class A   &  $0.14$ & $0.58$  & $0.11$  & $0.11$  & $78.3$  & $25.3$  \\
Class B   &  $0.12$ & $0.44$  &  $0.10$ & $0.10$  & $51.1$  &  $16.3$ \\
Class C   & $0.11$  & $0.36$  & $0.11$  &  $0.11$ & $26.3$  &  $10.2$ \\
\hline
\end{tabular}
\end{table}

From the OCs classified in class A, $139$ of them have radial velocity measurements available, with $85$ having more than two stars with radial velocity. For those, the mean dispersion of the radial velocities within cluster member stars is $5.47$ km$\cdot$s$^{-1}$. For the OCs in class B, $93$ from $236$ have radial velocity measurements, and $42$ counts with more than two stars with the measure. The mean radial velocity dispersion for class B clusters is $6.59$ km$\cdot$s$^{-1}$. Finally, for class C clusters, only $38$ have stars with radial velocity, of which $20$ have measurements for more than two stars. In this case, the mean dispersion is $11.81$ km$\cdot$s$^{-1}$. Part of these dispersions can be due to multiplicity. Since the clustering did not take into account the radial velocity in order to detect the OCs, this external check shows the degree of contaminant stars that clusters in each class may have.

\begin{figure}[!htb]
\includegraphics[width = 1.\columnwidth]{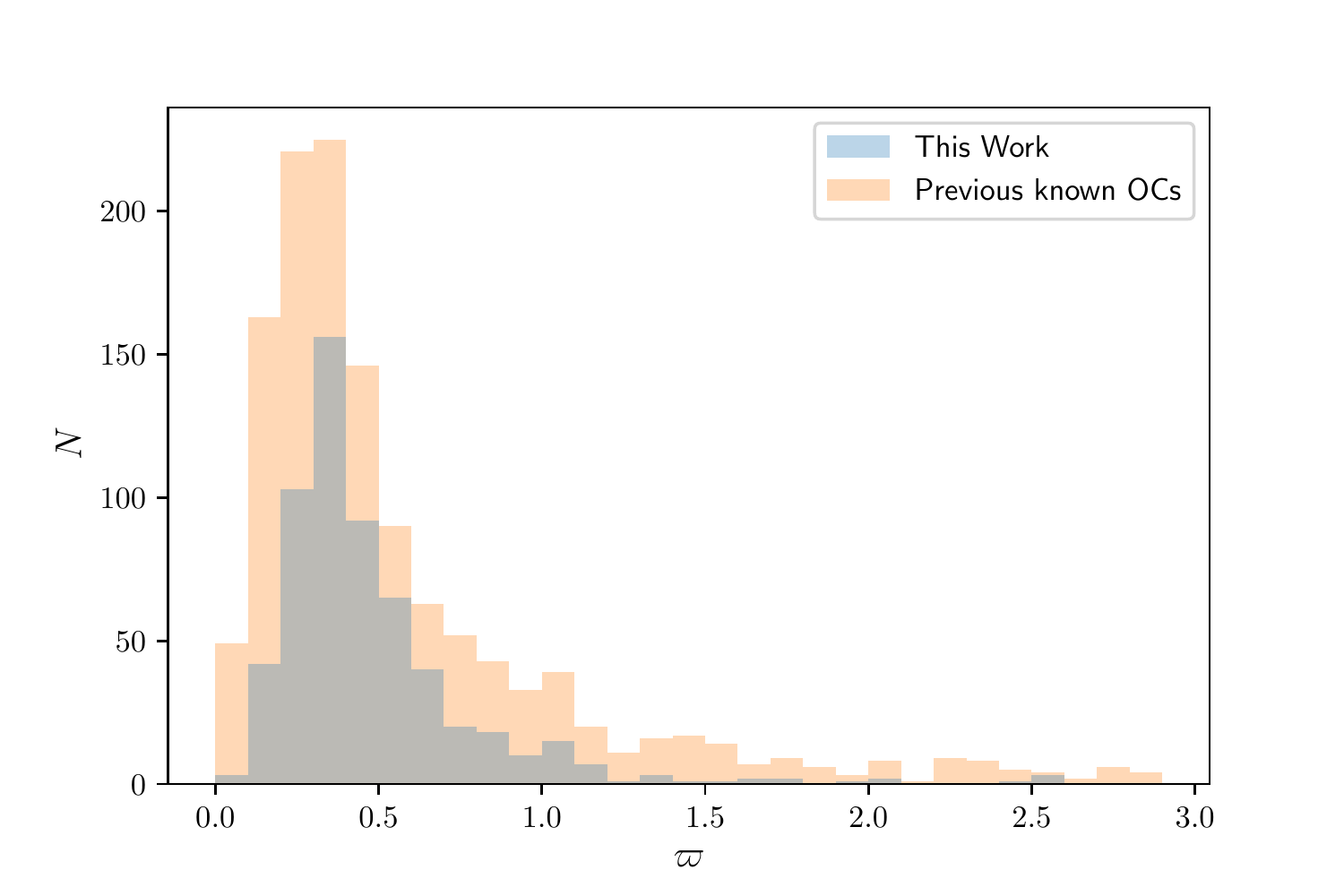}
\caption{Parallax histogram of the new OCs (light blue) and OCs known previous to this study (light orange), \textit{i.e.} CG18, CG19, \citet{tristan_catalogue} and \citet{coin_clusters}.}
\label{fig:par_hist}
\end{figure}

\subsubsection{Comments on the new OCs}
\label{subsubsec:general_comments}

The newly found clusters have mean parallaxes ranging from $0.09$ to $2.58$ mas. Estimating their distance as the inverse of their mean parallax yields distances from $387$ pc to $\sim 11$ kpc. Inverting parallaxes is however not a good approach for objects with large relative parallax uncertainties \citep{2018A&A...616A...9L}, and a more sophisticated method should be applied to estimate the distance to the most distant OCs. \figrefalt{fig:par_hist} shows a comparison between the distribution of parallaxes of the known OCs with the new findings, with light orange representing previous known OC and light blue representing OCs found in this paper. The OCs found represent an increase in the OC census of a $18\%$ in clusters closer than $1$ kpc, $54\%$ in clusters between $1 - 2$ kpc and $49\%$ in clusters further than $2$ kpc. 

\begin{figure}[!htb]
\includegraphics[width = 1.\columnwidth]{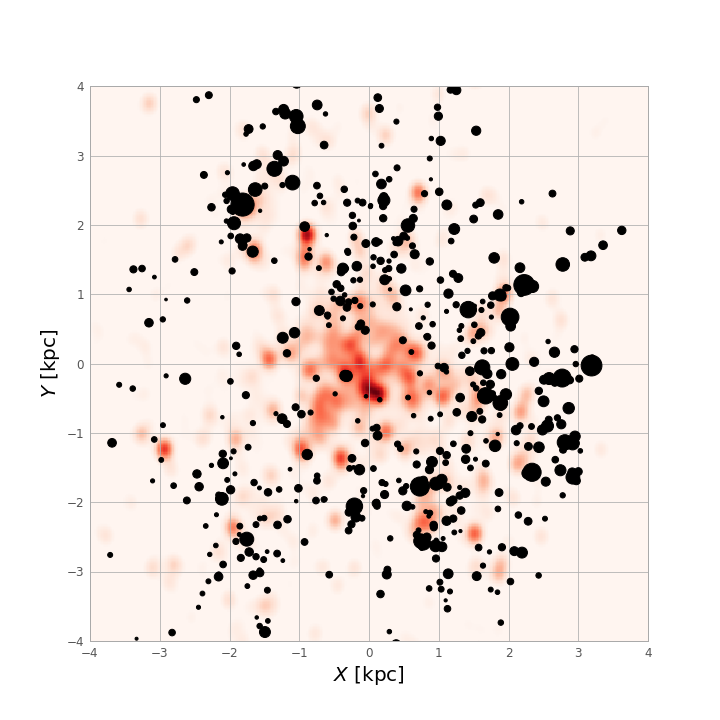}
\caption{Distribution of the OCs projected in the $X-Y$ plane. Previous known OCs \citet[CG18,CG19,][]{tristan_catalogue,coin_clusters} are shown as a density map in red. New OCs reported in this work are shown as black dots where the size is proportional to the number of members of each cluster.}
\label{fig:xyplane}
\end{figure}

\begin{figure*}[!htb]
\includegraphics[width = 1.\textwidth]{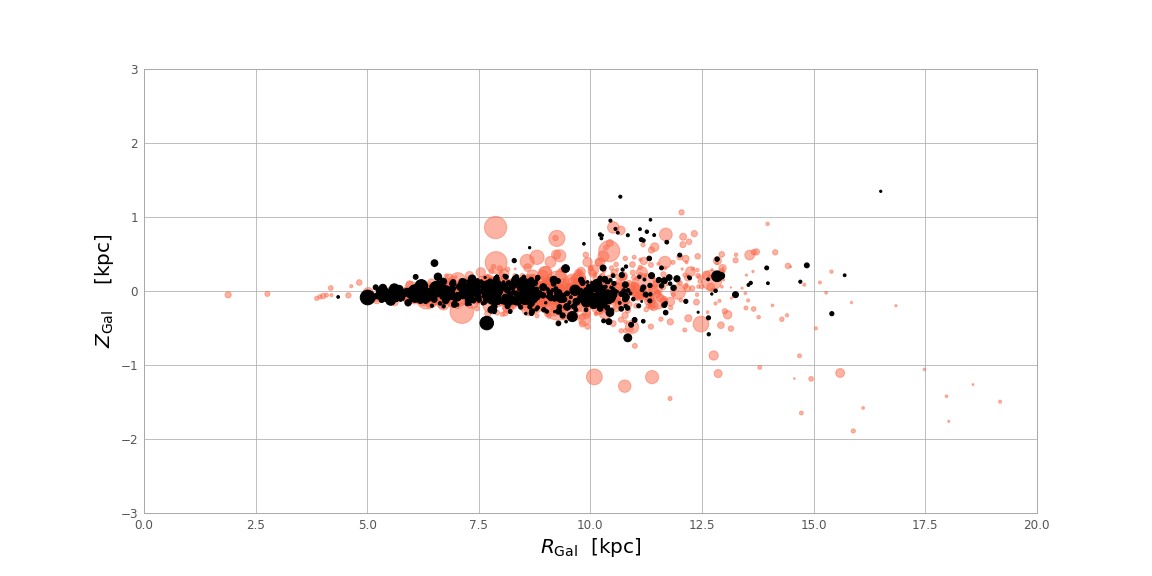}
\caption{Distribution of the OCs in $R-Z$ in Galacto-centric coordinates. Previous known OCs \citet[CG18,CG19,][]{tristan_catalogue,coin_clusters} are shown as red dots while new OCs are shown in black dots. For both, the dot size in proportional to the number of members of the cluster.}
\label{fig:RvsZ}
\end{figure*}

The distribution of the new OCs in the Galactic plane is shown in \figref{fig:lbplane} (projection in the $X-Y$ plane in \figref{fig:xyplane}). An $83.5\%$ of the new OCs are located at Galactic latitudes $|b| < \ang{5}$, for $\ang{5} < |b| < \ang{10}$ we find $8.2\%$ of OCs while only an $8.3\%$ is found at $|b| > \ang{10}$. The black dots represent the new found OCs (with its angular size proportional to the number of members) while the red density contours represent the known ones. We see that the distribution of the new OCs follow a similar distribution that the previous reported ones. In these figures we can see that the present study detected relatively few new objects between Galactic longitudes $\ang{140}$ and $\ang{210}$. This region has already been the target of two cluster searches using \textit{Gaia} DR2 data \citep[in CG19 and][]{coin_clusters}, and fewer objects are left to be discovered here. \figrefalt{fig:RvsZ} shows a distribution of the known (red dots) and new OCs (black dots), we see that none of the new OCs are found at high $|Z_{Gal}|$ in the inner disc ($R_{Gal} < 7$ kpc) where it is unlikely to find real OCs \citep{2019arXiv191107075C}.

\subsubsection{Specific remarks on UBC~$274$}
\label{subsubsec:UBC_2}

UBC~$274$ is a new OC found at a relatively low Galactic latitude $(b \sim \ang{-12.8})$ at a distance of $d \sim 2$ kpc. It is the clearest new detection made with our method, \textit{i.e.} the cluster most found within the pairs of $(L,minPts)$ explored, one of the more massive OC we can find (with $365$ stars), and one of the biggest in size. There are $15$ stars with radial velocity measurements, of which $13$ are in agreement with a mean value of $-22.92$ km$\cdot$s$^{-1}$, and a standard deviation of $1.26$ km$\cdot$s$^{-1}$, so they are compatible with the membership. The non-compatible stars have a radial velocity of $-10.68$ and $-8.00$ km$\cdot$s$^{-1}$, at $9\sigma$ and $11\sigma$ difference, respectively, they may be field stars or multiple stars.

\begin{figure*}[!htb]
\centering
\includegraphics[width = 1.\textwidth]{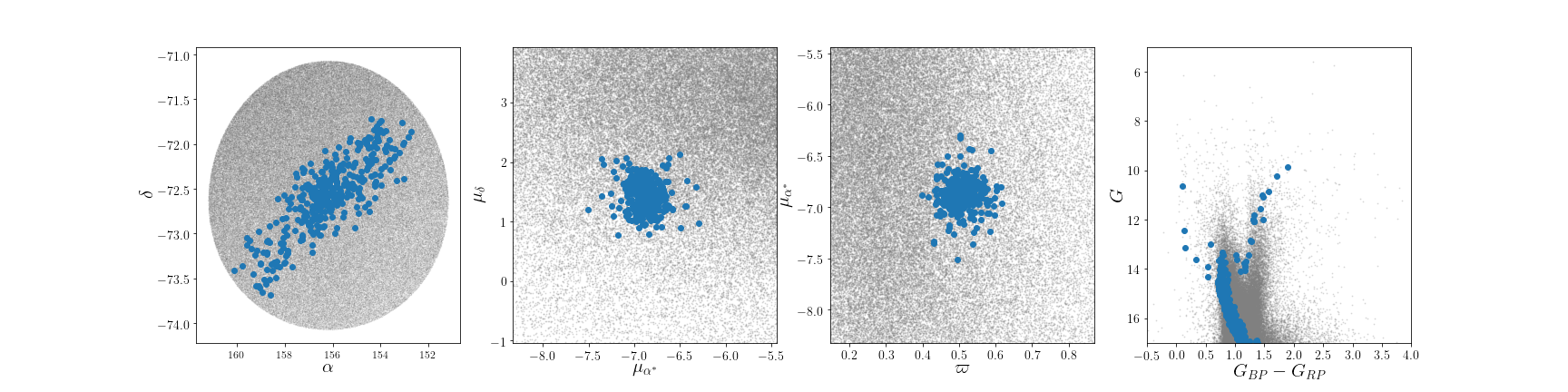}
\caption{Distribution of the member stars of UBC~$274$ (blue points) in comparison with field stars (grey points). The leftmost plot is a distribution in position $(\alpha,\delta)$. The inner left plot shows the proper motion vector diagram while the inner right plot includes the parallax $(\varpi,\mu_{\alpha^*})$. The rightmost plot is a CMD.}
\label{fig:cand_2}
\end{figure*}

\figrefalt{fig:cand_2} shows a distribution of its member stars in the five astrometric dimensions, and in a CMD. Their members show a concentrated clump in $(\varpi,\mu_{\alpha^*},\mu_\delta)$, well distinguishable from the field stars. It shows an elongated shape in the spatial distribution, in the direction of the proper motion. The CMD shows a clean isochrone from which we can estimate an age of $\sim 3$ Gyr. Less than $20\%$ of the previously known clusters have ages larger than $1$ Gyr, and only $5\%$ larger than $2$ Gyr. We can also identify some blue straggler candidates.

\begin{figure}[!htb]
\centering
\includegraphics[width = 1.\columnwidth]{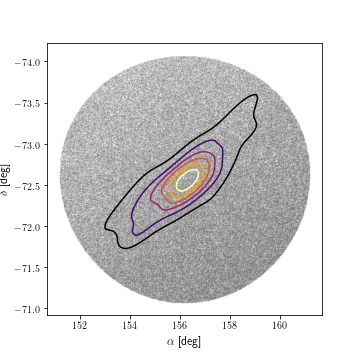}
\caption{Density contours for the members in cluster UBC~$274$, and field stars (grey points). UBC~$274$ shows an elongated shape in its outskirts.}
\label{fig:cand_2_contour}
\end{figure}

Tidal tails in intermediate and old age OCs due to disruption by the gravitational field have been detected in well known clusters like the Hyades, Praesepe or Coma Berenices by \citep{2019A&A...621L...2R,2019A&A...627A...4R,2019ApJ...877...12T} based on \textit{Gaia} DR2. The elongation of UBC~$274$ (\figref{fig:cand_2_contour}) suggests that it is another example of disruption taking place.

\subsubsection{Substructure in star forming regions}
\label{subsubsec:starforming}

\begin{figure}[!htb]
\centering
\includegraphics[width = 1.\columnwidth]{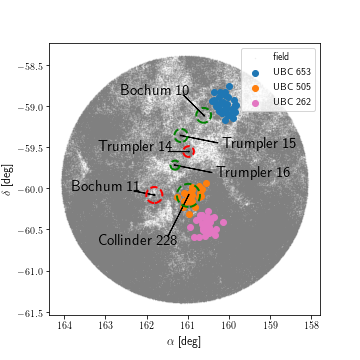}
\caption{Region around the Carina Nebula. Grey points represent field stars, while points in blue, orange or pink represent UBC~$653$, UBC~$505$ and UBC~$262$ respectively. The dashed circle represents locations of the OCs \citet{tristan_catalogue}, which are related to the Carina Nebula. Dashed green circles are objects found by our method and dashed red circles are objects not found.}
\label{fig:carina}
\end{figure}

It has been known for a long time that star forming regions are in groups and form structures and filaments \citep[\textit{e.g.}][]{2015A&A...584A..26B}. After \textit{Gaia} DR2 it has been possible to accurately distinguish their spatial and kinematic substructure in several star forming regions \citep{2018A&A...620A.172Z,2019MNRAS.490..440L,2019A&A...630A.137G,2019A&A...626A..17C} and even study the internal dynamics of these groups. We identified several objects possibly related to known star forming regions. For instance, in the Carina Nebula, we are able to find $7$ groups which are related to the nebula. \figrefalt{fig:carina} shows the spatial distribution of those groups. The points in different colours represent the stars found for each of the new UBC clusters, and dashed circles represent known clusters related to the nebula (in green the ones that our method finds, and in red the ones our method does not). We see that even in a blind search, we are able to detect several subgroups which can be related to the same structure. For instance, Collinder~$228$ and UBC~$505$ share sky coordinates but they are found as two different objects due to the difference in parallax, which is $0.42$ and $0.29$ mas, respectively.

\section{Conclusions}
\label{sec:conclusions}

We have devised a methodology to blindly search for open clusters in the Galactic disc, using the \textit{Gaia} DR2 astrometric and photometric data. The method is based on two ML algorithms, first an unsupervised learning algorithm (DBSCAN) detects overdensities in the astrometric space $(l,b,\varpi,\mu_{\alpha^*},\mu_\delta)$ and after that, a supervised ANN recognises the isochrone pattern that some of these statistical overdensities (the ones that correspond to real OCs) show in a CMD, identifying them as actual OCs.

In order to scan the whole Galactic disc using a strategy driven by the targeted OCs and not the computational limitations, the method has to be adapted to a Big Data environment. We use the PyCOMPSs parallelisation scheme to deploy the clustering algorithm to the MareNostrum Supercomputer, at the BSC. This enables the possibility to search for overdensities independently of the density of the region, \textit{e.g.} higher density regions such as the direction of the Galactic centre. Once the statistical densities are detected, and due to the large amount of them, a more reliable photometric confirmation of the candidate is needed. This is reached by applying deep learning methods to an ANN, which outperform the simple multi-layer perceptron when two dimensional correlations are present (a CMD in $G$ \textit{vs.} $G_{BP}-G_{RP}$).

The methodology is able, even in a blind search where the parameters are tuned to find the largest amount of OCs, to find substructures in richer regions or even features of individual objects such as its tidal tails. This suggests that with a fine tuning of the parameters, the methodology can be adapted to study single objects in more detail.

The method was first devised using TGAS data in CG18, and successfully applied to a low density disc region (the Galactic anticentre) using \textit{Gaia} DR2 in CG19, finding a total of $76$ new OCs. In this paper, the method is applied to the whole Galactic disc ($|b| < \ang{20}$) up to magnitude $G = 17$, finding a total of $582$ so far unknown OCs, which represents an increase of $45\%$ in the number of previously known OCs.

The OCs found represent an increase of $18\%$ up to $1$ kpc, $54\%$ between $1$ and $2$ kpc, and $49\%$ further than $2$ kpc. The mean angular size of the clusters found is $\ang{0.13}$ and the mean number of members is $58.3$. One of the most interesting clusters found is UBC~$274$ about $3$ Gyrs old at $b = \ang{-12.8}$, which shows an elongated shape due to the disruption by tidal tails.

\begin{acknowledgements}

ACG thanks Dr. T. Antoja for her comments on the writing; ACG also thanks Dr. Jordi Vitri\`a and Dr. Santi Segu\'i for their useful comments 
on the ANN implementation and training.

This work has made use of results from the European Space Agency (ESA)
space mission {\it Gaia}, the data from which were processed by the {\it Gaia
Data Processing and Analysis Consortium} (DPAC).  Funding for the DPAC
has been provided by national institutions, in particular the
institutions participating in the {\it Gaia} Multilateral Agreement. The
{\it Gaia} mission website is \url{http: //www.cosmos.esa.int/gaia}. The
authors are current or past members of the ESA {\it Gaia} mission team and
of the {\it Gaia} DPAC.

This work was partially supported by the MINECO (Spanish Ministry of Economy) through grant ESP2016-80079-C2-1-R and RTI2018-095076-B-C21 (MINECO/FEDER, UE), and MDM-2014-0369 of ICCUB (Unidad de Excelencia 'María de Maeztu'). 

This work has received funding from the European Union’s Horizon 2020 research and innovation programme under the Marie Skłodowska-Curie grant agreement H2020-MSCA-COFUND-2016-754433.
This work has been partially supported by the Spanish Government (SEV2015-0493), by the Spanish Ministry of Science and Innovation (contract TIN2015-65316-P), by Generalitat de Catalunya (contract 2014-SGR-1051).
The research leading to these results has also received funding from the collaboration between Fujitsu and BSC (Script Language Platform).

L.C. acknowledges support from "programme national de physique stellaire" (PNPS) and from the "programme national cosmologie et galaxies".

This research has made use of the TOPCAT \citep{topcat}.
This research has made use of the VizieR catalogue access tool, CDS,
Strasbourg, France. The original description of the VizieR service was
published in A$\&$AS 143, 23.
\end{acknowledgements}

\bibliographystyle{aa} 
\bibliography{bibliography}


\end{document}